\documentclass[twocolumn,aps,pra,superscriptaddress,footinbib,amsmath,amssymb,showpacs]{revtex4-1} %
\usepackage{graphicx}
\usepackage{bm}
\usepackage[colorlinks,	linkcolor=blue,anchorcolor=blue,citecolor=blue,bookmarksnumbered]{hyperref}
\usepackage{float}


\usepackage{babel}
\begin{document}
\title{Mechanical cooling at the bistable regime of a dissipative optomechanical
cavity with a Kerr medium}
\author{Ye Liu}
\affiliation{Fujian Key Laboratory of Quantum Information and Quantum Optics and
Department of Physics, Fuzhou University, Fuzhou 350116, China}
\author{Yang Liu}
\affiliation{Fujian Key Laboratory of Quantum Information and Quantum Optics and
Department of Physics, Fuzhou University, Fuzhou 350116, China}
\author{Chang-Sheng Hu}
\affiliation{Department of Physics, Anhui Normal University, Wuhu 241000, China}
\author{Yun-Kun Jiang}
\affiliation{Fujian Key Laboratory of Quantum Information and Quantum Optics and
Department of Physics, Fuzhou University, Fuzhou 350116, China}
\author{Huaizhi Wu}
\affiliation{Fujian Key Laboratory of Quantum Information and Quantum Optics and
Department of Physics, Fuzhou University, Fuzhou 350116, China}
\author{Yong Li}
\affiliation{Center for Theoretical Physics, Hainan University,
Haikou 570228, China}

\begin{abstract}
In this paper, we study static bistability and mechanical cooling
of a dissipative optomechanical cavity filled with a Kerr medium.
The system exhibits optical bistability for a wide input-power range
with the power threshold being greatly reduced, in contrast to the
case of purely dissipative coupling. At the bistable regime, the membrane
can be effectively cooled down to a few millikelvin from the room
temperature under the unresolved sideband condition, where the effective
mechanical temperature is a nonmonotonic function of intracavity intensity
and reaches its minimum near the turning point of the upper stable
branch. When the system is in the cryogenics environment, the effective
mechanical temperature at the bistable regime shows a similar feature
as in the room temperature case, but the optimal cooling appears at
the monostable regime and approaches the mechanical ground state.
Our results are of interest for further understanding bistable optomechanical
systems, which have many applications in nonclassical state preparations
and quantum information processing. 
\end{abstract}
\maketitle

\section{INTRODUCTION}

Cavity optomechanical systems, which study the interplay between
light and mechanical motion, have important applications in fundamental
tests of quantum mechanics, precision measurement, and quantum information
processing \citep{Aspelmeyer2014}. For a generic setup, the radiation
pressure force exerted by the light field typically induces a displacement-dependent
cavity frequency, and thus enables a dispersive coupling between the
optical and mechanical degrees of freedom.  Since the radiation-pressure
coupling is intrinsically nonlinear, optomechanical systems can exhibit
different types of nonlinear behaviors, depending on the input power
and the detuning of the driving laser with respect to the cavity resonance.
In the blue-detuned regime, a strong driving can trigger rich nonlinear
phenomena, such as dynamical multistability \citep{2006FMDynamicalMultistability,2008Metzger,2011HeinrichComptesRendus,Wu2013,Walter2014,Mercade2021},
instability \citep{2008Ludwig}, synchronization \citep{Heinrich2010Collective-dyna,Weissa,2013BagheriTangSynchronization,Walter2014},
and chaotic motion \citep{Carmon2007,Larson2011,Zhang2022,Bakemeier2014}.
In the red-detuned regime, one obtains the static optical bistability,
where the mechanical mode in the low-temperature limit acts as a Kerr
nonlinearity for the cavity mode \citep{Aldana2013}.  The red-detuned
regime is also considered as the appropriate regime for ground-state
cooling of the mechanical motion \citep{Marquardt2007a,Wilson-Rae2007,Vanner2013,Teufel:2011uq,Sawadsky2015b,Painter,OaConnell:2010rt},
which is typically a prerequisite for nonclassical state preparation
\citep{2012QianFMNonclassicalStates,Vanner2013a,Liao2016,Milburn2016,Li2018,Shomroni2020},
quantum information processing \citep{2013Palomakilehnertentanglement,Riedinger2018,Marinkovi2018,Fiaschi2021},
and quantum-limited measurements \citep{Stannigel2010,Chang2014,Nielsen2002,Stannigel2011,Rips2013,Tagantsev2019}
with mechanical oscillators. With dispersive optomechanical coupling,
sideband cooling of mechanical oscillation into its ground state has been
experimentally demonstrated in the red-detuned regime out of the optical
bistable region \citep{Painter,Teufel:2011uq,Meenehan2015}. Besides,
there are \textcolor{black}{a} few works which have also looked into
the relationship between optical bistability and quantum effects.
It has been shown that squeezing \citep{Fabre1994,Kronwald2014a,Agarwal2016,Wollman2015a,Hu2018,Aggarwal2020a,Kustura2022}
and light-mechanical entanglement \citep{Ghobadi2011a,Ghobadi2011,Hu2020}
induced by optomechanical interactions become maximal for the parameter
regime close to the threshold of the bistability or instability, and
particularly,  the entanglement is counter-intuitively not a monotonic
function of the optomechanical coupling strength in the bistable region
\citep{Ghobadi2011a}.

There exists another kind of cavity optomechanics \citep{2009ElsteQuNoiseInterference,2010AgarwalEIT,2010WeisKippenbergOMIT,2011XuerebDissipativeOptomechanics,Weiss2013},
where the cavity linewidth depends on the mechanical displacement,
giving rise to a dissipative coupling between the mechanical and the
optical degrees of freedom. Dissipative optomechanical coupling can
be realized with superconducting microwave circuits \citep{2009ElsteQuNoiseInterference}
or with a Michelson--Sagnac interferometer containing a semitransparent
movable membrane \citep{2011XuerebDissipativeOptomechanics,Tarabrin2013},
and moreover, it has been experimentally demonstrated with a microdisc
resonator coupled to a nanomechanical waveguide \citep{Li2009}, with
a photonic crystal split-beam nanocavity \citep{Wu2014a}, and with
graphene drums coupled to a high-Q microsphere \citep{Cole2015}.
 In analogy to dispersive optomechanics, a dissipative coupling
also allows for mechanical cooling \citep{2009ElsteQuNoiseInterference,David2015,Zhang2015c,Zhang2019a},
nonclassical state preparation \citep{Teufel2011,Gu2013,Yan2015a,Qu2015,Tagantsev2018,Huang2020},
quantum-limited position measurements \citep{Hryciw2015,Tagantsev2019},
and quantum sensing (of force \citep{Huang2017,Qamar2018,Mehmood2020,He2022}
and speed \citep{Vyatchanin2016,Ashour2020}).  In contrast to purely
dispersive systems, one can observe negative-damping instability \citep{2009ElsteQuNoiseInterference,Weiss2013,Tarabrin2013,Sawadsky2015b}
and self-sustained oscillations \citep{Huang2018d} unconventionally
for weak cavity driving of red detunings by involving a dissipative
coupling. Although a purely dissipative coupling also has a nonlinear
effect on the intracavity intensity, which is expected to show static
bistability in principle \citep{Weiss2013}, however, we note that
the system is normally hard to run into bistability for typical experimental parameters
with low input power \citep{Sawadsky2015b}. Therefore, nonclassical
properties of a dissipative optomechanical setup at the bistable regime
may only be studied by introducing an extra nonlinearity or by considering
a hybrid optomechanical system \citep{Kyriienko2014,Pelka2021}. 

In this paper, we study optical bistability and mechanical cooling
at the bistable regime of a dissipative optomechanical system, which
is implemented with a Michelson--Sagnac interferometer containing
a movable membrane and a Kerr medium. We consider that the mechanical
motion only causes a shift of the cavity damping rate, and does not
vary the cavity frequency, corresponding to a purely dissipative coupling.
As a clear advantage over the setting without Kerr nonlinearity,
the cavity intensity can exhibit optical bistability at regular laser
driving power of tens of milliWatt (mW), where the input-power-dependent
cavity intensity displays a characteristic $S$-shaped curve and part
of the upper branch of the curve turns unstable due to optomechanical
coupling. In this regard, we note that the bistable region exists
for a broad power range. In contrast, the optical bistability is sensitive
to power fluctuation in typical dispersive systems \citep{Ghobadi2011a,Aldana2013},
and is inaccessible by the purely dissipative system with the Michelson--Sagnac
interferometer and the input power of $\sim$100 mW. We then study
mechanical cooling at the bistable regime,   and find that the steady-state
mechanical temperature is a non-monotonic function of intracavity
intensity (or optomechanical coupling). The dissipative coupling allows
the mechanical membrane to be effectively cooled from room temperature
down to a few millikelvin in the unresolved sideband regime, and the optimal cooling condition
appears at the upper branch close to the turning point. We
further show that the membrane initially in a cryogenic environment
of 0.1 K has similarly bistable features, and can be cooled down close to the ground state in the unresolved sideband regime for the case of only branch.
Our findings are of interest for further studying the optomechanical
nonclassical properties in the presence of static bistability.

The paper is organized as follows: Section II introduces the dissipative
optomechanical system and describes the linearization of the equations
of motion around the steady state. We also show the stability conditions
required to satisfy in this framework. Section III introduces the
effective mechanical susceptibility, noise spectrum and the effective
temperature of the mechanical membrane. Section IV shows how bistability
arises in the red-detuned regime, and the dependence of the photon
number on the driving power and detuning, which leads us to a discussion
of cooling on both stable branches in the bistable regime shown in
Section V. Section VI is a further discussion and conclusion.

\section{model, linearization of the Hamiltonian, and the stability condition}

\begin{figure}
\includegraphics[width=1\columnwidth]{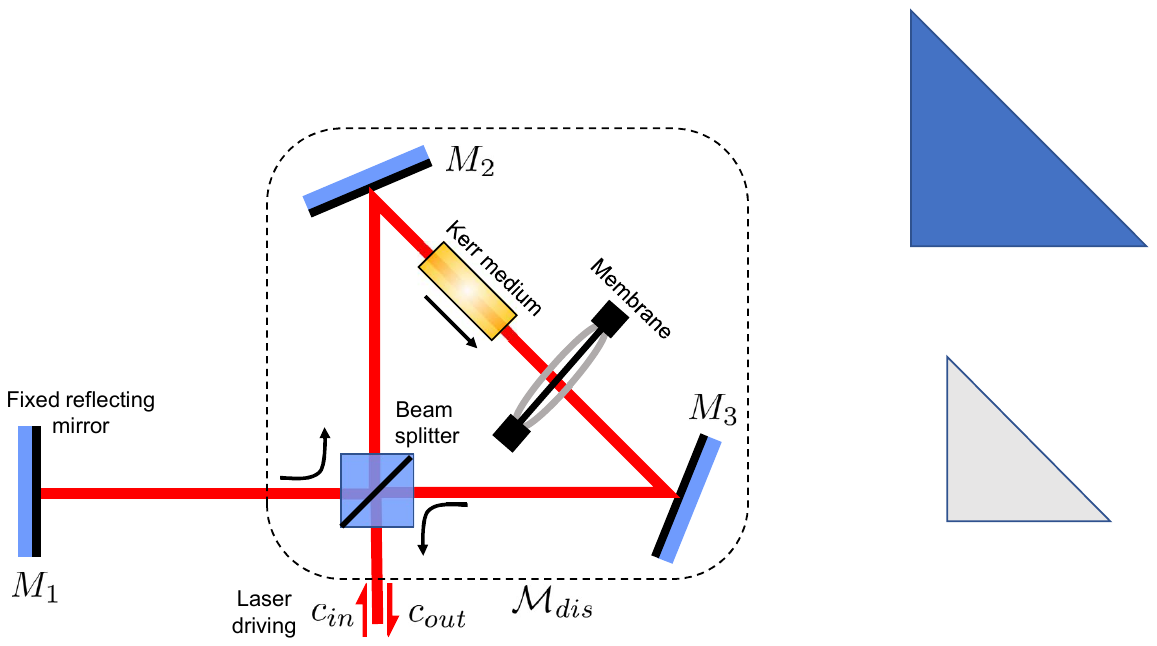}

\caption{\label{Scheme}\textcolor{black}{(Color online) Schematics of the
dissipative optomechanical setup. We consider a Michelson-Sagnac interferometer,
which consists of three fixed perfect reflecting mirrors $M_{i}$
($i=1,2,3$) and a fixed beam splitter (BS). A movable membrane and
a Kerr medium are placed in the middle of $M_{2}$ and $M_{3}$ \citep{2011XuerebDissipativeOptomechanics}.
A strong classical driving field is input to the interferometer via
the vertical ports of the BS. The part encircled by dashed line can
be regarded as an effective end mirror $\mathcal{M}_{dis}$. The linewidth
of the cavity depends on the membrane displacement, which leads to
an effective dissipative coupling between the cavity mode and the
mechanical motion. }}
\end{figure}

As shown in Fig. \ref{Scheme}, we consider an optomechanical Michelson-Sagnac
interferometer \citep{2011XuerebDissipativeOptomechanics}, which
includes a mechanical membrane (of mass $m$, frequency $\omega_{m}$,
and damping rate $\gamma_{m}$), and a Kerr medium (of the nonlinearity
strength $U$) placed along the light propagation path. When an external
driving field (with frequency $\omega_{l}$ and power $\mathcal{P}$)
is injected into the interferometer at the beam splitter, the system
can be effectively described as a dissipative optomechanical cavity,
where the resonance frequency $\omega_{c}(q)$ and decay rate $\kappa(q)$
depend on the displacement of the mechanical membrane $q$ \citep{2011XuerebDissipativeOptomechanics}.
The Hamiltonian of the system in a frame rotating at the input laser
frequency $\omega_{l}$ is given by \citep{2011XuerebDissipativeOptomechanics}
\begin{eqnarray}
H & = & \hbar[\omega_{c}(q)-\omega_{l}]c^{\dagger}c+\frac{1}{2}(m\omega_{m}^{2}q^{2}+\frac{p^{2}}{m})+i\hbar\sqrt{2\kappa(q)}\nonumber \\
 &  & \times[c^{\dagger}(\epsilon_{l}+c_{in})-c(\epsilon_{l}+c_{in}^{\dagger})]-\hbar Uc^{\dagger2}c^{2},
\end{eqnarray}
where $c$ $(c^{\dagger})$ is the annihilation (creation) operator
of the cavity field satisfying the commutation relation $[c,c^{\dagger}]=1$,
and $q$ and $p$ are the mechanical displacement and momentum operators
with $[q,p]=i\hbar$. $\epsilon_{l}=\sqrt{\mathcal{P}/\hbar\omega_{l}}$
is the laser driving strength (assumed to be real for simplicity)
and $c_{in}$ is the input vacuum noise satisfying the usual nonvanishing
correlation function $\langle c_{in}(t)c_{in}^{\dagger}(t^{\prime})\rangle=\delta(t-t^{\prime}).$
Typically, the mechanical displacement only weakly modulates $\omega_{c}(q)$
and $\kappa(q)$ such that we can expand them to just the linear order
of $q$, i.e. \citep{2009ElsteQuNoiseInterference,2010AgarwalEIT,2010WeisKippenbergOMIT,2011XuerebDissipativeOptomechanics,Weiss2013}
\[
\omega_{c}(q)=\omega_{c}+g_{\omega}q,\text{ }\kappa(q)=\kappa+g_{\kappa}q,
\]
where $g_{\omega}=\partial\omega_{c}(q)/\partial q$ and $g_{\kappa}=\partial\kappa(q)/\partial q$
are dispersive and dissipative coupling constants between the cavity
field and the membrane, respectively. Moreover, it has been shown
that the dispersive coupling constant $g_{\omega}$ can be set to
zero if the complex reflectivity and transmissivity of the beam splitter
are appropriately selected \citep{2011XuerebDissipativeOptomechanics}.
As a result, only the cavity decay rate depends on the mechanical
displacement, and the setup is referred to as a (purely) dissipative
optomechanical system. For the purely dissipative regime $g_{\omega}=0$,
the Hamiltonian of the system, in terms of the rescaled mechanical
position and momentum operators $Q=q/\sqrt{2}x_{\text{zpf}}$ and
$P=p(\sqrt{2}x_{\text{zpf}}/\hbar)$, can be rewritten by 
\begin{eqnarray}
H & = & \hbar\Delta c^{\dagger}c+\frac{1}{2}\hbar\omega_{m}(Q^{2}+P^{2})-\hbar Uc^{\dagger2}c^{2}\nonumber \\
 &  & +i\hbar\sqrt{2\kappa}(1+\frac{g}{2\kappa}Q)[c^{\dagger}(\epsilon_{l}+c_{in})-c(\epsilon_{l}+c_{in}^{\dagger})],
\end{eqnarray}
where $\Delta=\omega_{c}-\omega_{l}$ is the laser detuning with respect
to the cavity resonant frequency, and $g=\sqrt{2}g_{\kappa}x_{\text{zpf}}$
is the rescaled dissipative coupling constant. 

From the above Hamiltonian, we can derive the detailed dynamics of
the system via the standard Langevin equations
\begin{eqnarray}
\dot{Q} & = & \omega_{m}P,\nonumber \\
\dot{P} & = & -\omega_{m}Q-i\frac{g}{\sqrt{2\kappa}}[c^{\dagger}(\epsilon_{l}+c_{in})-c(\epsilon_{l}+c_{in}^{\dagger})]\nonumber \\
 &  & -\gamma_{m}P+\xi,\nonumber \\
\dot{c} & = & -(\kappa+gQ+i\Delta)c+\sqrt{2\kappa}(1+\frac{g}{2\kappa}Q)(\epsilon_{l}+c_{in})\nonumber \\
 &  & +2iUc^{\dagger}c^{2},\label{eq:Lang_Eqs}
\end{eqnarray}
where the mechanical thermal noise $\xi$ is zero mean valued and
fulfills the two-time correlation function  $\langle\xi(t)\xi(t^{\prime})\rangle=\frac{1}{2\pi}\frac{\gamma_{m}}{\omega_{m}}\text{\ensuremath{\int}}\omega e^{-i\omega(t-t^\prime)}[1+{\rm coth}(\frac{\hbar\omega}{2k_{B}T})]d\omega$ \citep{2008GenesGroundStateCooling,Giovannetti2001},
with $T$ being the thermal temperature of the environment. By denoting
$\bar{O}=\langle O\rangle$ as the steady-state value of $O=Q,P,c$,
and using the fact $\langle c_{in}\rangle=0$ and $\langle\xi\rangle=0$,
one can obtain the steady-state semiclassical solutions of $\bar{O}$
by solving Eq. (\ref{eq:Lang_Eqs}) with $\langle\dot{O}\rangle=0$,
which gives rise to
\begin{equation}
\bar{Q}=ig\sqrt{\mathcal{P}_{l}}\frac{(\bar{c}-\bar{c}^{*})}{\omega_{m}},\text{ }\bar{P}=0,\label{eq:Q_ss}
\end{equation}
\begin{equation}
\bar{c}=\frac{(\kappa+\tilde{\kappa})\sqrt{\mathcal{P}_{l}}}{\tilde{\kappa}+i(\Delta-2U\vert\bar{c}\vert^{2})},\label{eq: C_ss}
\end{equation}
with $\tilde{\kappa}=\kappa+g\bar{Q}$, and $\mathcal{P}_{l}=\epsilon_{l}^{2}/2\kappa$.
Eq. (\ref{eq: C_ss}) implies that a static bistability may occur
even though the Kerr nonlinearity is set to zero \citep{Weiss2013}.
However, we find that the system cannot run into bistability for the
typical experimental parameters with the input power of a few hundred
mW \citep{Sawadsky2015b}. Thus, the Kerr nonlinearity is of great
essential for the bistable regime focused on in this work.

\begin{figure}
\includegraphics[width=0.9\columnwidth]{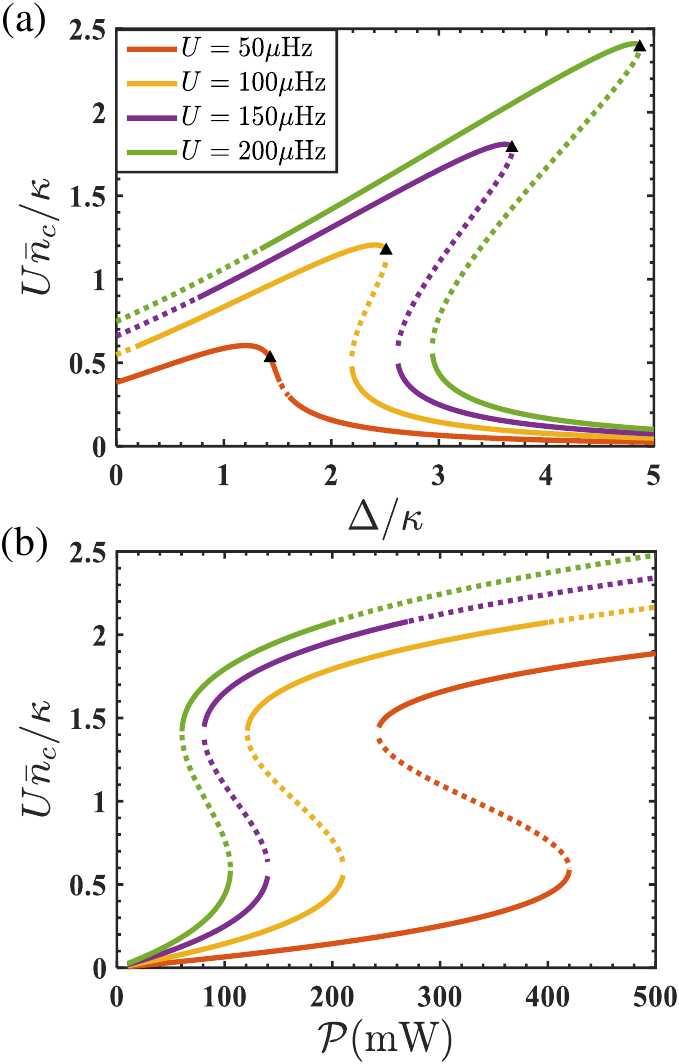}\caption{\label{Fig_Classical_Bistability}\textcolor{black}{(Color online) The normalized cavity intensity $U\bar{n}_{c}/\kappa$ (a) as a function
of the dimensionless detuning $\Delta/\kappa$ for driving power
$\mathcal{P}=100$ mW, and (b) as a function of the driving power
$\mathcal{P}$ for ${\color{red}{\color{black}\Delta/\kappa=3}}$,
with the nonlinearity strengths being $U=50$ $\mu$Hz (orange), $100$
$\mu$Hz (yellow), $150$ $\mu$Hz (purple) and $200$ $\mu$Hz (green),
respectively. We consider the set of typical experimental parameters
\citep{Sawadsky2015b}: the wavelength of the input laser $\lambda=2\pi c/\omega_{l}=1064$
nm, the cavity decay rate $\kappa=2\pi\times1.5$ MHz, the dissipative
optomechanical coupling rate $g=2\pi\times0.1$ Hz, the mechanical
frequency $\omega_{m}=2\pi\times136$ kHz, and the mechanical damping
rate $\gamma_{m}=2\pi\times0.23$ Hz. The cavity intensity displays
a characteristic $S$-shaped curve, with the stable and unstable parts
being indicated by the solid and dotted lines, respectively. The black triangles mark the parameter conditions plotted in Fig. \ref{high frequency and damping rate}. }}
\end{figure}

Assuming that the mean photon number in the cavity is far more than
1 (i.e. $\langle c^{\dagger}c\rangle\gg1$), and the static stability
conditions (shown later) are met, we proceed with the usual linearization
around steady state by decomposing each observable as the sum of its
steady-state mean value and a small quantum fluctuation: $O=\bar{O}+\delta O$.
Neglecting higher order terms for the fluctuations, we obtain
\begin{eqnarray}
\delta\dot{Q} & = & \omega_{m}\delta P,\nonumber \\
\delta\dot{P} & = & -\omega_{m}\delta Q+i\frac{g\epsilon_{l}}{\sqrt{2\kappa}}(\delta c-\delta c^{\dagger})-\gamma_{m}\delta P\nonumber \\
 &  & -i\frac{g}{\sqrt{2\kappa}}(\bar{c}^{*}c_{in}-\bar{c}c_{in}^{\dagger})+\xi,\nonumber \\
\delta\dot{c} & = & -(\tilde{\kappa}+i\tilde{\Delta})\delta c+g\zeta\delta Q+2i\tilde{U}e^{i\phi}\delta c^{\dagger}\nonumber \\
 &  & +\frac{1}{\sqrt{2\kappa}}(\kappa+\tilde{\kappa})c_{in},\label{Eq:QLangevin_EOM}
\end{eqnarray}
where $\tilde{U}=U|\bar{c}|^{2}$ , $\tilde{\Delta}=\Delta-4\tilde{U}$,
$\phi=\text{arg}(\bar{c}^{2})$ and $\zeta=\sqrt{\mathcal{P}_{l}}-\bar{c}$.
The Kerr nonlinearity introduces two effects to the system: First,
the cavity field $\delta c$ can be squeezed via the parametric Hamiltonian
$-\hbar\tilde{U}e^{i\phi}\delta c^{\dagger2}+\text{H.c.}$; second,
the cavity frequency is effectively shifted by $4\tilde{U}$, which
is essential for stabilization of the system to achieve mechanical
cooling. To examine the dynamic stability of the system, we further
rewrite the Langevin equations by $\dot{u}(t)=Mu(t)+n(t)$, which
are expressed in terms of the quadrature operators $u(t)=[\delta Q,\delta P,\delta x,\delta y]^{T}$
with $\delta x=\frac{1}{\sqrt{2}}(\delta c+\delta c^{\dagger})$ and
$\delta y=\frac{1}{\sqrt{2}i}(\delta c-\delta c^{\dagger})$ being
the amplitude and phase quadratures of the cavity field, $n(t)=[0,\xi+\frac{g}{\sqrt{2\kappa}}(-\bar{c}_{i}x_{in}+\bar{c}_{r}y_{in}),\frac{\kappa+\tilde{\kappa}}{\sqrt{2\kappa}}x_{in},\frac{\kappa+\tilde{\kappa}}{\sqrt{2\kappa}}y_{in}]^{T}$
with $x_{{\rm in}}=\frac{1}{\sqrt{2}}(c_{{\rm in}}+c_{{\rm in}}^{\dagger})$
and $y_{{\rm in}}=\frac{1}{\sqrt{2}i}(c_{{\rm in}}-c_{{\rm in}}^{\dagger})$
being the input vacuum noises, and the evolution matrix
\begin{widetext}
\begin{equation}
M=\left(\begin{array}{cccc}
0 & \omega_{m} & 0 & 0\\
-\omega_{m} & -\gamma_{m} & 0 & -g\sqrt{2\mathcal{P}_{l}}\\
\sqrt{2}g\left(\sqrt{\mathcal{P}_{l}}-\bar{c}_{r}\right) & 0 & -\tilde{\kappa}-2\tilde{U}{\rm sin}\phi & \tilde{\Delta}+2\tilde{U}{\rm cos}\phi\\
-\sqrt{2}g\bar{c}_{i} & 0 & -\tilde{\Delta}+2\tilde{U}{\rm cos}\phi & -\tilde{\kappa}+2\tilde{U}{\rm sin}\phi
\end{array}\right)
\end{equation}
\end{widetext}
with $\bar{c}_{r}=\frac{1}{2}(\bar{c}+\bar{c}^{*}),$ $\bar{c}_{i}=\frac{1}{2i}(\bar{c}-\bar{c}^{*})$.
Followed by a consideration of the Routh-Hurwitz criterion \citep{DeJesus1987}, the real
part of the eigenvalues of $M$ should be strictly negative such that
the system is stable, which gives rise to the stability conditions
below: 
\begin{widetext}
\begin{equation}
s_{1}=2\tilde{\kappa}[(\tilde{\kappa}+\gamma_{m})^{2}+\tilde{\Delta}^{2}-4\tilde{U}^{2}]+2\sqrt{\mathcal{P}_{l}}g^{2}\bar{c}_{i}\omega_{m}+\gamma_{m}\omega_{m}^{2}>0,\label{eq:stable_s1}
\end{equation}
\begin{eqnarray}
s_{2} & = & 2\sqrt{\mathcal{P}_{l}}g^{2}[(\bar{c}_{r}-\sqrt{\mathcal{P}_{l}})(\tilde{\Delta}-2\tilde{U}\text{cos}\phi)-\bar{c}_{i}(\tilde{\kappa}+2\tilde{U}\text{sin}\phi)]+(\tilde{\kappa}^{2}+\tilde{\Delta}^{2}-4\tilde{U}^{2})\omega_{m}>0,\label{eq:stable_s2}
\end{eqnarray}
\begin{eqnarray}
s_{3} & = & s_{1}[(\tilde{\kappa}^{2}+\tilde{\Delta}^{2}-4\tilde{U}^{2})\gamma_{m}+2\omega_{m}(\tilde{\kappa}\omega_{m}-\sqrt{\mathcal{P}_{l}}g^{2}\bar{c}_{i})]-\left(2\tilde{\kappa}+\gamma_{m}\right){}^{2}s_{2}\omega_{m}>0.\label{eq:stable_s3}
\end{eqnarray}
\end{widetext}
Moreover, since the studies of the current work focus on the bistable
regime, we have to also confirm that the fluctuation of photon number
is much less than the classical mean value, i.e. $\left(\langle c^{\dagger}c\rangle-|\bar{c}|^{2}\right)/|\bar{c}|^{2}\ll1$,
such that our discussions stay within the range of validity of the
linearization approximation.

\section[Optical bistability]{Optical bistability}

According to Eq. (\ref{eq: C_ss}), we obtain a third-order polynomial
root equation for the mean-field cavity occupation.
\begin{equation}
4U^{2}\bar{n}_{c}^{3}-4\Delta U\bar{n}_{c}^{2}+\left(\tilde{\kappa}^{2}+\Delta^{2}\right)\bar{n}_{c}=\mathcal{P}_{l}(\kappa+\tilde{\kappa})^{2},\label{eq:polynormial}
\end{equation}
where $\bar{n}_{c}=|\bar{c}|^{2}$. Equation (\ref{eq:polynormial})
indicates that the steady-state cavity intensity $\bar{n}_{c}$ can
have either one or three solutions, depending on the number of real
roots of the polynomial. By considering the parameter regime, where the mechanical frequency and damping rate are $(\omega_{m}, \gamma_{m})/\kappa= (0.091,1.53\times10^{-7})$, the optomechanical coupling strength is $g/\kappa=6.67\times10^{-8}$, the cavity driving strength is $\epsilon_{l}^{2}/\kappa=5.68\times10^{10}$, in Fig. \ref{Fig_Classical_Bistability}(a),
we plot the rescaled mean-field occupation $U\bar{n}_{c}/\kappa$
as a function of the detuning $\Delta/\kappa$ for $\kappa=2\pi\times1.5$ MHz and then the driving
power $\mathcal{P}=\epsilon_{l}^{2}(\hbar\omega_{l})=100$ mW. As the Kerr nonlinearity increases from
$U=50$ $\mu$Hz ($U/\kappa=0.53\times~10^{-11}$) to $U=200$ $\mu$Hz ($U/\kappa=2.12\times10^{-11}$), the cavity line shape, which
is approximately Lorentzian for $U=0$, becomes more and more asymmetric
and tilts until. The system is stable only when the stability criteria
Eqs. (\ref{eq:stable_s1})-(\ref{eq:stable_s3}) are obeyed. In general,
the violation of the criterion $s_{1}>0$ and $s_{2}>0$ always yields
an unstable middle branch, while the additional criterion for the
optomechanical system $s_{3}>0$ can turn part of the upper or only
branch unstable, see the tails near the resonance $\Delta/\kappa=0$.

Since part of the upper branch of the $S$-shaped curves may turn unstable and the system can be in the monostable regime for $\Delta/\kappa\rightarrow 0$ with the parameters under consideration,
in Fig. \ref{Fig_Classical_Bistability}(b), we plot the rescaled
mean-field occupation $U\bar{n}_{c}/\kappa$ as a function of the
driving power $\mathcal{P}$ for the fixed detuning $\Delta/\kappa=3$.
For an increasing driving power $\mathcal{P}$, in all cases the mean-field
occupations $\bar{n}_{c}$ have three branches, which form a characteristic
$S$-shaped curve. As discussed before, the violation of $\{s_{1},s_{2}\}>0$
again gives rise to an unstable middle branch, and the upper branch
is stable only in a finite segment corresponding to $s_{3}>0$. Note that for a large detuning $\Delta/\kappa\gg3$ (and for the nonlinearity strengths under consideration), it requires a far stronger driving to exhibit optical bistability, and meanwhile the laser detuning itself must remain far less than the free spectral range of the cavity. 

Without optomechanical coupling, the Kerr nonlinearity can lead to
optical bistability with a fully stable upper branch \citep{Aldana2013};
with purely dissipative optomechanical coupling but without the Kerr
nonlinearity, the bistable behavior can only turn up at the strong
laser driving power of a few Watts, which is normally hard to access
for typical quantum optical experiments. Here, by combining the Kerr
nonlinearity with the dissipative coupling interaction, we are able
to observe optical bistability at the driving power on the order of
$\sim100$ mW, and moreover, the optical bistability is much less
insensitive to power fluctuation, in contrast to that in dispersive
optomechancal systems with typical parameters (as shown by Aldana
et al. \citep{Aldana2013}), which is confined in a power range of
a few mW due to the limit of the stability condition $s_{3}$, and
therefore is sensitive to power disturbance.

 Recalling that the resonant frequency of the cavity can be shifted by $\sim U\bar{n}_{c}$ because of the Kerr nonlinearity, here we have carefully examined the effective detunings $\tilde{\Delta}=\Delta-4U\bar{n}_{c}$, which is in the range of $-5<\tilde{\Delta}/\kappa<5$ with respect to the $S$-shaped curves in Fig. \ref{Fig_Classical_Bistability}(b), and confirmed that both $\Delta$ and $\tilde{\Delta}$ remain far less than the free spectral range of the cavity, which is given by $\text{FSR}=c/(2L)=1.7\text{ GHz}\sim180\kappa$ with the effective cavity length $L=0.087$ m and the cavity decay rate $\kappa/2\pi\sim1.5$ MHz  \citep{Sawadsky2015b}.

\section{Effective mechanical susceptibility, noise spectrum, and final temperature }

\begin{figure*}
\centering{}\includegraphics[width=0.7\textwidth]{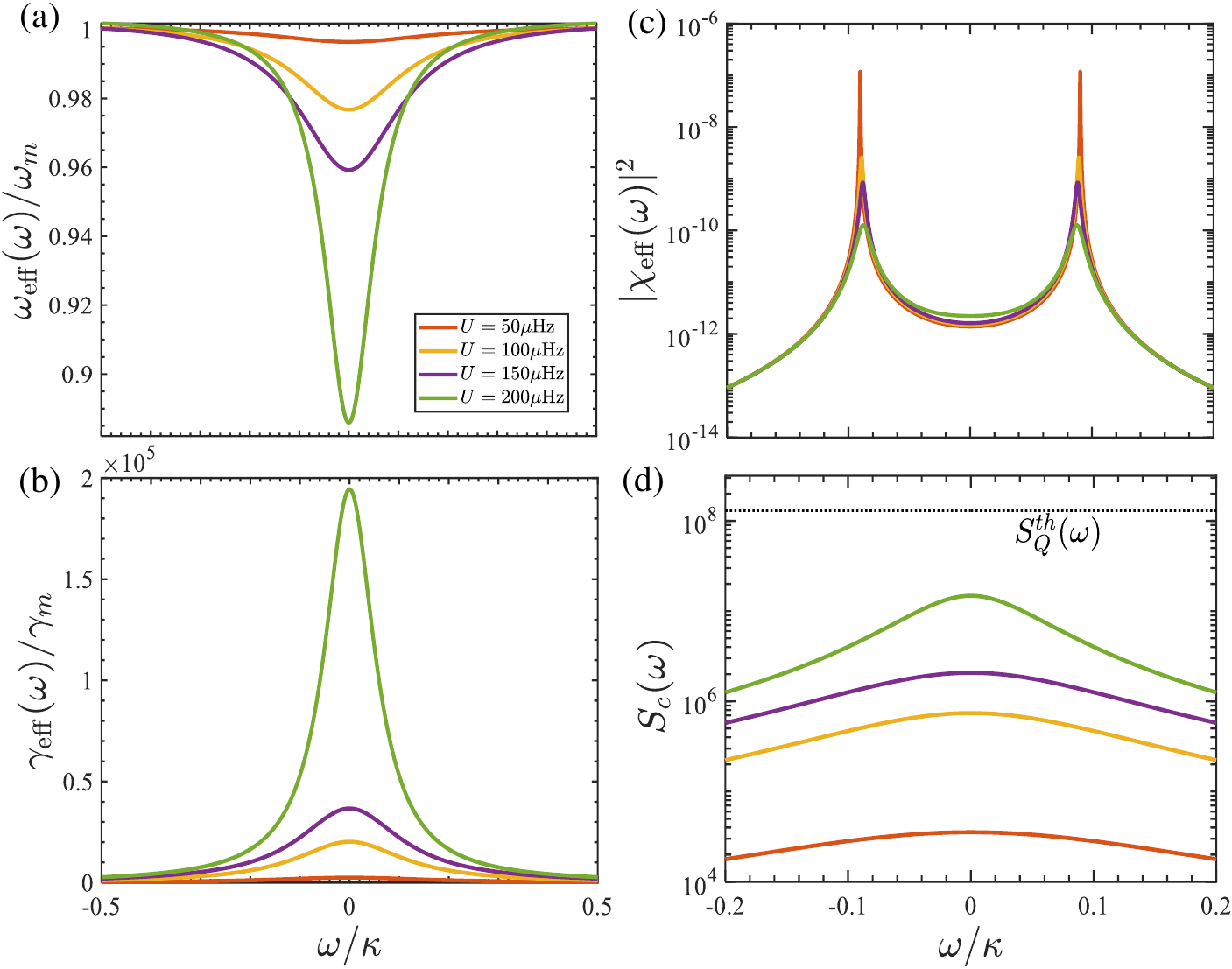}\caption{\label{high frequency and damping rate}\textcolor{black}{(Color online)
(a) The normalized effective resonance frequency $\omega_{{\rm eff}}(\omega)/\omega_{m}$,
(b) the normalized effective damping rate $\gamma_{{\rm eff}}(\omega)/\gamma_{m}$,
(c) the modulus of effective susceptibility $|\chi_{{\rm eff}}(\omega)|^{2}$,
and (d) spectrum of the input vacuum noise as a function of the normalized
frequency $\text{\ensuremath{\omega/\kappa}}$. The dotted line in
(d) indicates the mechanical thermal noise, which for the environment
of the room temperature $T=293$ K considered here is approximately
frequency independent. The Kerr nonlinearity and the cavity detuning
for the curves {[}from top to bottom in (a){]} are $U=50$ $\mu$Hz
and $\Delta=1.43\kappa$ (orange), $U=100$ $\mu$Hz and $\Delta=2.51\kappa$
(yellow), $U=150$ $\mu$Hz and $\Delta=3.68\kappa$ (purple), $U=200$
$\mu$Hz and $\Delta=4.87\kappa$ (green), respectively, }
corresponding to the parameters indicated by the black triangles in Fig. \ref{Fig_Classical_Bistability}(a). Other
parameters are the same as in Fig. \ref{Fig_Classical_Bistability}.}
\end{figure*}

In order to gain more insights into the dynamics of the system and
the limitations for effective cooling of the membrane, we analytically
derive the effective mechanical susceptibility and the spectra of
the input noises. We first Fourier transform Eq. (\ref{Eq:QLangevin_EOM})
by using $\delta O(\omega)=\int_{-\infty}^{+\infty}\delta O(t)e^{i\omega t}d\omega$,
after simple algebra, then obtain the expression of the position fluctuations
in the frequency domain,
\begin{equation}
\delta Q(\omega)=\chi_{{\rm eff}}(\omega)\left[\Lambda(\omega)c_{{\rm in}}(\omega)+\Lambda^{*}(-\omega)c_{{\rm in}}^{\dagger}(\omega)+\xi(\omega)\right],\label{eq:dQ_noises}
\end{equation}
where
\begin{equation}
\chi_{{\rm eff}}^{-1}(\omega)=\chi_{{\rm m}}^{-1}(\omega)+\Xi(\omega)\label{eq:chi_eff1}
\end{equation}
is the effective mechanical susceptibility with 
\begin{equation}
\chi_{{\rm m}}^{-1}(\omega)=\frac{\omega_{m}^{2}-i\omega\gamma_{m}-\omega^{2}}{\omega_{m}},
\end{equation}
\begin{equation}
\Xi(\omega)=-i\sqrt{\mathcal{P}_{l}}g^{2}\frac{\tilde{\chi}_{c}^{-1}(\omega)\zeta-\left[\tilde{\chi}_{c}^{-1}(-\omega)\right]^{*}\zeta^{*}}{\chi_{c}^{-1}(\omega)\left[\chi_{c}^{-1}(-\omega)\right]^{*}-4\tilde{U}^{2}},
\end{equation}
\begin{equation}
\chi_{c}^{-1}(\omega)=\tilde{\kappa}-i(\omega+\tilde{\Delta}),
\end{equation}
\begin{equation}
\tilde{\chi}_{c}^{-1}(\omega)=\chi_{c}^{-1}(\omega)+2i\tilde{U}e^{-i\phi},
\end{equation}
and 
\begin{equation}
\Lambda(\omega)=i\frac{g}{\sqrt{2\kappa}}\left[\frac{\sqrt{\mathcal{P}_{l}}(\kappa+\tilde{\kappa})\tilde{\chi}_{c}^{-1}(\omega)}{\chi_{c}^{-1}(\omega)\left[\chi_{c}^{-1}(-\omega)\right]^{*}-4\tilde{U}^{2}}-\bar{c}^{*}\right].
\end{equation}
\noindent Eq. (\ref{eq:dQ_noises}) shows that the mechanical fluctuations
are given by the product of the effective susceptibility $\chi_{{\rm eff}}(\omega)$
and the sum of fluctuations arising from two uncorrelated noise terms,
namely the back action force noise of the cavity mode and the mechanical
Brownian noise. Moreover, the effects of the dissipative optomechanical
coupling $g$ and the Kerr nonlinearity $U$ have been collected into
the quantity $\Xi(\omega)$, which may be called as the optomechanical
self-energy \citep{2006FMDynamicalMultistability}. Since the dissipative
coupling is weak ($g/\kappa\ll1$), the effective susceptibility $\chi_{{\rm eff}}(\omega)$
will have a single resonance, whose property is just modified by
the presence of the Kerr nonlinearity $U$. To see the physical insight,
we then rewrite the effective mechanical susceptibility as
\begin{equation}
\chi_{{\rm eff}}(\omega)=\frac{\omega_{m}}{\omega_{{\rm eff}}^{2}-\omega^{2}-i\omega\gamma_{{\rm eff}}(\omega)},\label{chi_eff}
\end{equation}
where 
\begin{equation}
\omega_{{\rm eff}}^{2}(\omega)-\omega_{m}^{2}=-\frac{g^{2}\sqrt{\mathcal{P}_{l}}\omega_{m}\mu}{\lambda^{2}(\omega)+4\omega^{2}\tilde{\kappa}^{2}}
\end{equation}
and
\begin{equation}
\gamma_{{\rm eff}}(\omega)-\gamma_{m}=\frac{g^{2}\sqrt{\mathcal{P}_{l}}\omega_{m}\nu}{\lambda^{2}(\omega)+4\omega^{2}\tilde{\kappa}^{2}}, \label{eq:gamma_opt}
\end{equation}

\noindent with
\[
\lambda(\omega)=\widetilde{\kappa}^{2}+\tilde{\Delta}^{2}-\omega^{2}-4\tilde{U}^{2},
\]
\[
\mu=-2{\rm Im}\left[\tilde{\chi}_{c}^{-1}(0)\zeta\right]\lambda(\omega)+4\omega^{2}\widetilde{\kappa}\bar{c}_{i},
\]
\[
\nu=-2\lambda(\omega)\bar{c}_{i}-4\widetilde{\kappa}{\rm Im}\left[\tilde{\chi}_{c}^{-1}(0)\zeta\right].
\]
By comparing Eq. (\ref{eq:chi_eff1}) with Eq. (\ref{chi_eff}),
one can figure out the meaning of both the imaginary and the real
parts of $\Xi(\omega)$, which evaluated at the original resonance
frequency $\omega=\omega_{m}$ are a shift of the mechanical frequency
(``optical spring\textquotedblright ) and some optomechanical damping
rate, respectively \citep{Wilson-Rae2007,Marquardt2007a}.

Furthermore, by considering the nonzero correlation functions $\langle c_{{\rm in}}(\omega)c_{{\rm in}}^{\dagger}(\omega^{\prime})\rangle=2\pi\delta(\omega-\omega^{\prime})$
and $\langle\xi(\omega)\xi(\omega^{\prime})\rangle=2\pi\frac{\gamma_{m}}{\omega_{m}}\omega[1+{\rm coth}(\frac{\hbar\omega}{2k_{B}T})]\delta(\omega-\omega^{\prime})$
for the noise operators, we obtain the spectrum of the mechanical
position and momentum \citep{Huang2020}, 
\begin{eqnarray}
S_{Q}(\omega) & = & |\chi_{{\rm eff}}(\omega)|^{2}\left[S_{c}(\omega)+S_{Q}^{th}(\omega)\right],\label{eq:spectrum_Q}
\end{eqnarray}
\begin{equation}
S_{P}(\omega)=\frac{\omega^{2}}{\omega_{m}^{2}}S_{Q}(\omega),
\end{equation}
\noindent where $S_{c}(\omega)=\frac{1}{2}\left[\Lambda(\omega)\Lambda^{*}(\omega)+\Lambda(-\omega)\Lambda^{*}(-\omega)\right]$
and $S_{Q}^{th}(\omega)=\frac{\gamma_{m}}{\omega_{m}}\omega{\rm coth}(\frac{\hbar\omega}{2k_{B}T})$
are spectra of the input vacuum noise and the mechanical thermal noise,
respectively. Integrating $S_{Q}(\omega)$ and $S_{P}(\omega)$ over
all the frequency range then gives rise to the variance of the mechanical
position and momentum: 
\begin{eqnarray}
\langle\delta Q^{2}\rangle & = & \frac{1}{2\pi}\int_{-\infty}^{+\infty}d\omega S_{Q}(\omega),\\
\langle\delta P^{2}\rangle & = & \frac{1}{2\pi}\int_{-\infty}^{+\infty}d\omega S_{P}(\omega).\label{eq:variance}
\end{eqnarray}

\noindent As a figure of merit for cooling, we calculate the final
occupation number of the mechanical membrane via
\begin{eqnarray}
n_{m} & = & \frac{1}{2}[\langle\delta Q^{2}\rangle+\langle\delta P^{2}\rangle-1],\label{eq:phonon_no}
\end{eqnarray}
and consider the effective temperature in the membrane with
\begin{eqnarray}
T_{{\rm eff}} & = & \frac{\hbar\omega_{m}}{k_{B}{\rm ln}(1+\frac{1}{n_{m}})}.
\end{eqnarray}
In addition, by following the same recipe as from Eq. (\ref{eq:spectrum_Q})
to Eq. (\ref{eq:phonon_no}), we can further calculate the variance
of the amplitude and phase quadratures of the cavity field, which
enables us to examine optical squeezing and the fluctuation of the
photon number above its classical mean value $\bar{n}_{c}$ via $\delta n_{c}=\frac{1}{2}[\langle\delta x^{2}\rangle+\langle\delta y^{2}\rangle-1]$.
The latter is then used to confirm the validity of the linearization
approximation, which is a prerequisite for the discussion of mechanical
cooling at the bistable regime.

\begin{figure*}
\includegraphics[width=0.8\textwidth]{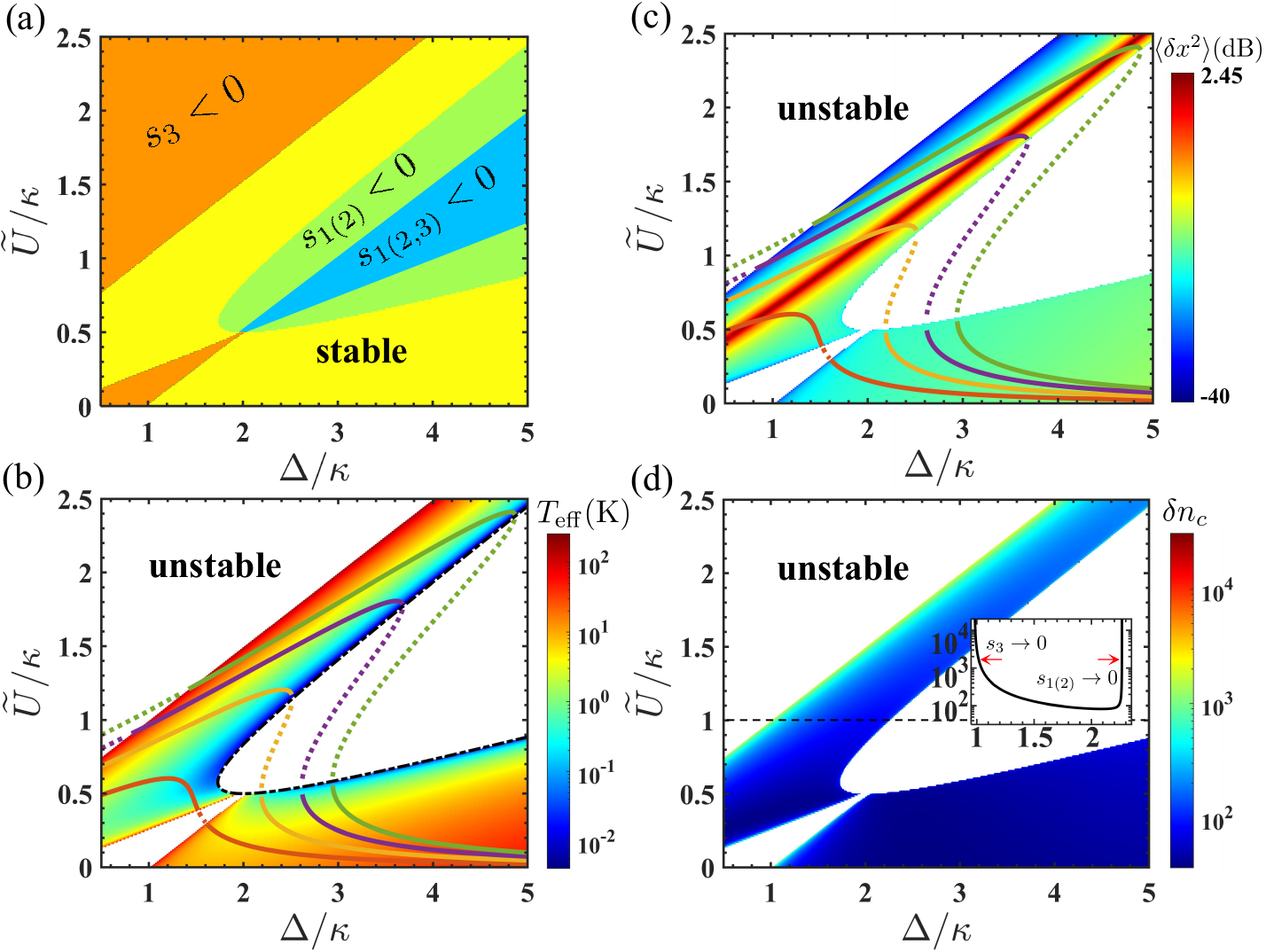}\caption{\label{T_eff_293}\textcolor{black}{(Color online) (a) Phase diagram
of the stable regime (yellow) confirmed via the Routh-Hurwitz criterion,
which requires $s_{1}$, $s_{2}$ and $s_{3}$ to be all positive.
The unstable region in green corresponds to $s_{1}<0$ (or $s_{2}<0$),
and the orange region corresponds to $s_{3}<0$. In the blue region,
$s_{1}$, $s_{2}$, and $s_{3}$ are all negative. (b) Effective temperature
of the membrane in the steady state versus $\Delta/\kappa$ and $\tilde{U}/\kappa$
by cooling from the initial room temperature $T=293$ K. The black dash-dotted line indicates $\lambda(\omega_m)=0$, which fits approximately with the boundary of the unstable region with $s_{1(2)}<0$, corresponding to the green zone in (a). (c) Variance
of the optical quadrature fluctuation $\langle\delta x^{2}\rangle$
and (d) the photon number fluctuation $\delta n_{c}$ versus $\Delta/\kappa$
and $\tilde{U}/\kappa$. The inset shows $\delta n_{c}$ versus $\Delta/\kappa$
for $\tilde{U}/\kappa=1$, and the photon number fluctuation diverges
at the stability boundary with $s_{j}\rightarrow0$, $j=1,2,3$. In
(b)-(d), the blank regions correspond to the classical unstable regime.
The four curves {[}in (b), (c){]}, and other parameters are the same
as those in Fig. \ref{Fig_Classical_Bistability}(a).}}
\end{figure*}

\section{Mechanical cooling in the bistable region}

We now focus on the mechanical cooling in the bistable regime or in
the vicinity of the unstable part for the case of only branch. In
the bistable regime, the fluctuation $\delta n_{c}$ in cavity intensity
around the steady state solution $\bar{n}_{c}$ may diverge as one
approaches the end of each stable branch. In order to stay within
the range of validity of the linearization approximation, we have
confirmed $\delta n_{c}/\bar{n}_{c}\ll1$ for the results shown below,
see further discussions later. In Figs. \ref{high frequency and damping rate}(a)
and \ref{high frequency and damping rate}(b), we show the normalized
effective resonance frequency $\omega_{{\rm eff}}(\omega)/\omega_{m}$
and the normalized effective damping rate $\gamma_{{\rm eff}}(\omega)/\gamma_{m}$
of the membrane as functions of the normalized frequency $\text{\ensuremath{\omega/\kappa}}$
for different strengths of the Kerr nonlinearity $U=\{50,100,150,200\}$
$\mu$Hz, where the corresponding cavity detunings are $\Delta/\kappa=1.43$,
2.51, 3.68, and 4.87, respectively. It can be seen that the overall
profiles of $\omega_{{\rm eff}}(\omega)/\omega_{m}$ for all values
of $U$ are above 0.88. As the Kerr nonlinearity increases, the optomechanically
induced frequency shift $\omega_{{\rm eff}}(0)/\omega_{m}$, and correspondingly,
the modified damping rate $\gamma_{{\rm eff}}(0)/\gamma_{m}$ gets
larger. Moreover, the responses of $\omega_{{\rm eff}}(\omega)/\omega_{m}$
to optomechanical interaction around the mechanical resonance (i.e.,
${\normalcolor {\color{red}{\color{black}\omega=\omega_{m}\sim0.091\kappa}}}$)
can be larger than $0.96$, with the modified damping rate $\gamma_{{\rm eff}}(\omega)$
increasing to $\sim2\pi\times0.02$ MHz for $U=200$ $\mu$Hz. It
implies that the optomechanically induced cooling of the membrane
can be realized with a relatively small ``optical spring'' frequency
shift for the set of Kerr nonlinearity $U=\{50,100,150,200\}$ $\mu$Hz under consideration. 

\textit{Room temperature} - We first suppose that the membrane is
in the room temperature environment with $T=293$ K, which corresponds
to the thermal phonon number of the membrane $n_{m}=4.49\times10^{7}$
according to $n_{m}=\left(e^{\hbar\omega_{m}/k_{B}T}-1\right){}^{-1}$.
In this case, the achievable final phonon number by optomechanical
cooling is limited by the input vacuum noise $c_{{\rm in}}(\omega)$
and the mechanical thermal noise $\xi(\omega)$, which make effects
via the response function (i.e., the modulus of the effective susceptibility)
$|\chi_{{\rm eff}}(\omega)|^{2}$, see Eqs. (\ref{eq:dQ_noises})
and (\ref{eq:spectrum_Q}). As shown in Fig. \ref{high frequency and damping rate}(c),
$|\chi_{{\rm eff}}(\omega)|^{2}$ displays two peaks at $\omega=\pm\omega_{{\rm eff}}(\omega_{m})$,
and the full width at half maximum (FWHM) of the peaks is determined
by $\gamma_{{\rm eff}}(\omega)$. The response function $|\chi_{{\rm eff}}(\omega)|^{2}$
ensures that the noise spectrum is only significant around a narrow
bandwidth centered about $\pm\omega_{{\rm eff}}(\omega_{m})$. Furthermore,
the spectrum $S_{c}(\omega)$ of the input vacuum noise is maximized
at the cavity resonance $\omega=0$, and the spectrum $S_{Q}^{th}(\omega)$
of the thermal noise at the thermal temperature $T=293$ K is almost
flat in the frequency range $\omega/\kappa\in(-0.2,0.2)$, see Fig.
\ref{high frequency and damping rate}(d). By comparison, the thermal
noise at room temperature dominates over the vacuum noise from the
cavity input. As a result, the position spectrum $S_{Q}(\omega)$
of the membrane can be approximately given by
\begin{equation}
S_{Q}(\omega)\approx|\chi_{{\rm eff}}(\omega)|^{2}\left(\frac{2k_{B}T}{\hbar\omega_{m}}+1\right)\gamma_{m},
\end{equation}
which is the response function $|\chi_{{\rm eff}}(\omega)|^{2}$ multiplied by a constant factor of the damping rate $(2k_{B}T/{\hbar\omega_{m}}+1)\gamma_{m}$ induced by the bath with temperature $T$. Thus, the position spectrum $S_{Q}(\omega)$ displays two peaks at $\omega=\pm\omega_{{\rm eff}}(\omega_{m})$ as well, and holds the linewidths $\gamma_{{\rm eff}}(\omega_{m})$ similar to $|\chi_{{\rm eff}}(\omega)|^{2}$.
When the strength of the Kerr nonlinearity increases {[}with the parameters being the same as in Fig. \ref{high frequency and damping rate}(a){]}, the peak values of $|\chi_{{\rm eff}}(\omega)|^{2}$ and therefore $S_{Q}(\omega)$ decrease sharply, while the corresponding FWHMs 
broaden, see Fig. \ref{high frequency and damping rate}(c).

Considering the room temperature condition and the unresolved sideband
limit, the final phonon number is then given by
\begin{equation}
n_{m}\approx\frac{k_{B}T\gamma_{m}}{2\pi\hbar\omega_{m}}\int_{-\infty}^{+\infty}d\omega\left(1+\frac{\omega^{2}}{\omega_{m}^{2}}\right)|\chi_{{\rm eff}}(\omega)|^{2}-\frac{1}{2}.
\end{equation}
Before we present the cooling effect, in Fig. \ref{T_eff_293}(a),
we first show the phase diagram of the stable regimes (defined by
the Routh-Hurwitz criterion) as a function of the dimensionless detuning
$\Delta/\kappa$ and the effective optical gain $\tilde{U}/\kappa$.
Previously, we have shown that the violation of the conditions $s_{j}>0$
($j=1,2,3$) has related to the stability of the upper or middle branch
when optical bistability occurs. Now, the cooling results are particularly
interesting because the effective mechanical temperature $T_{{\rm eff}}$
also strongly relies on the specific condition being violated, as
shown in Fig. \ref{T_eff_293}(b). For further insights, we have also
shown $\tilde{U}$ as a function of $\Delta/\kappa$ with the Kerr
nonlinearity strengths being $U=\{50,100,150,200\}$ $\mu$Hz, namely
the $S$-shaped curves shown in Fig. \ref{Fig_Classical_Bistability}(a).
For the cases of optical bistability, it is worth noting that $T_{{\rm eff}}$
gradually goes down as $s_{1}$ or $s_{2}$ approaches zero and suddenly
surges near $s_{1(2)}\approx0$ accompanied by divergence of $\delta n_{c}$
(see discussion later); $T_{{\rm eff}}$ reaches less than 5 mK in
the vicinity of the turning point at the upper branch. 
Indeed, this can be understood with the optomechanical damping rate $(\gamma_{{\rm eff}}(\ensuremath{\omega})-\gamma_{m})$ [as given by Eq. (\ref{eq:gamma_opt})], which can lead to extra damping and mechanical cooling when $(\gamma_{{\rm eff}}(\ensuremath{\omega})-\gamma_{m})$ is positive \citep{Marquardt2007a,Wilson-Rae2007}. Recalling that $S_{Q}(\omega)$ is approximately given by the product of $|\chi_{{\rm eff}}(\omega)|^{2}$ and the constant damping rate $\left(\frac{2k_{B}T}{\hbar\omega_{m}}+1\right)\gamma_{m}$, then for the frequency shift $\left(\omega_{\text{eff}}-\omega_{m}\right)$ being small and $\gamma_{{\rm eff}}\ll\kappa$, one can approach the best cooling effect by minimizing $|\chi_{{\rm eff}}(\omega)|^{2}\approx\gamma_{{\rm eff}}^{-2}(\ensuremath{\omega})$ [or maximizing $\left(\gamma_{{\rm eff}}(\ensuremath{\omega})-\gamma_{m}\right)]$  with $\omega=\omega_{m}$. Thus, we set $\lambda(\omega_{m})=0$ to minimize the denominator of Eq. (21), and find that $\tilde{\kappa}^{2}+\tilde{\Delta}^{2}-\omega_{m}^{2}-4\tilde{U}^{2}=0$ fits approximately with the boundary of the unstable region corresponding to $s_{1(2)}<0$ [as indicated by the black dash-dotted line in Fig. 4(b)], and moreover $\gamma_{{\rm eff}}(\ensuremath{\omega}_{m})$ is large near the boundary.
In contrast, for the tails of the curves close to $\Delta=0$ and for the case of only branch, as $s_{3}$ approaches
zero (i.e. the violation of the stability condition $s_{3}$), we find that $\lambda^{2}(\omega_{m})\gg4\omega_{m}^{2}\tilde{\kappa}^{2}$ and $v(\omega_{m})\rightarrow0$, and thus $T_{{\rm eff}}$ quickly goes up due to a vanishing optomechanical damping rate $(\gamma_{{\rm eff}}(\ensuremath{\omega_m})-\gamma_{m})$. Here, the linearization approximation is not justified and mechanical lasing instability may occur \citep{2006FMDynamicalMultistability,2011HeinrichComptesRendus,Wu2013}.
The lowest temperature of the membrane reached by combining the dissipative
coupling and the Kerr interaction is about three orders of magnitude
lower than that without the Kerr medium (i.e. $U=0$).

The Kerr nonlinearity can bring about not only the optical bistability,
but also a parametric amplification. To see that, we show the variance
of the $x$ quadrature of the cavity field in Fig. \ref{T_eff_293}(c).
We find that the intracavity squeezing of about 2.43 dB can be generated
around $\Delta\approx2\tilde{U}$ in both the monostable and the bistable
regime. For the cases of $U=50$ $\mu$Hz and $U=100$ $\mu$Hz, the
intracavity squeezing appears in the monostable region around $\Delta/\kappa=0.78$
and $\Delta/\kappa=2.15$, and the steady-state mechanical temperatures
are 1.38 K and 0.77 K, respectively. For $U=150$ $\mu$Hz and $U=200$
$\mu$Hz, the intracavity squeezing appears in the bistable region
and is only related to the upper branch, the steady-state mechanical
temperatures are 0.52 K and 0.4 K with $\Delta/\kappa=3.43$ and $\Delta/\kappa=4.68$,
respectively. Although the mechanical temperature can also reach the
same level $\sim$$0.1$ K at the lower branch, the variance in the
$x$-quadrature (or the $y$-quadrature) does not approach the zero-point
level anymore. On the other hand, as mentioned before, $\delta n_{c}\ll\bar{n}_{c}$
should hold for the effectiveness of the linearized dynamics, however,
as $s_{j}$ ($j=1,2,3$) approaches zero the variance in the $x$-quadrature
(or $y$-quadrature) and the photon number fluctuation $\delta n_{c}$
diverge, then the linearization approximation is not justified anymore.
As indicated in Fig. \ref{T_eff_293}(d), $\delta n_{c}$ is a few
tens or a few hundreds of quanta in most of the stable regions, and
becomes larger than $10^{3}$ only at the boundary with $s_{j}\rightarrow0$.
Moreover, the inset shows that $\delta n_{c}$ increases more sharply
at the turning point with ($s_{1(2)}\rightarrow0$), near the optimal
cooling regime.

\begin{figure}
\includegraphics[width=1\columnwidth]{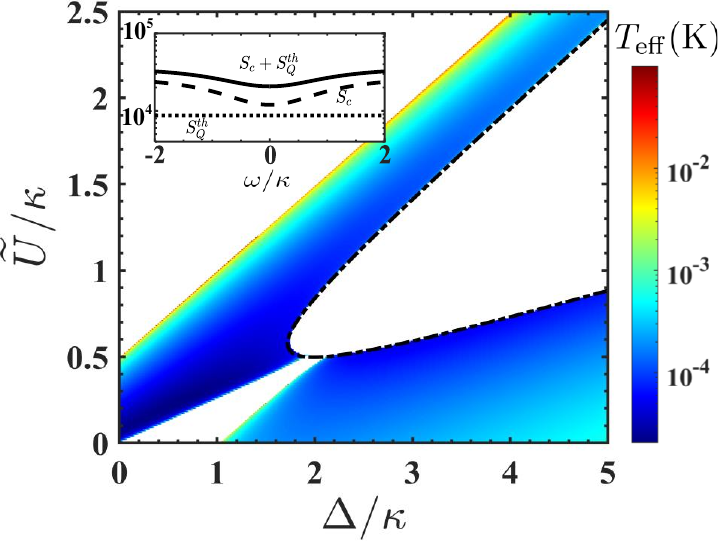}

\caption{\label{fig:cryogenics}\textcolor{black}{(Color online) Effective
temperature of the membrane in the steady state versus $\Delta/\kappa$
and $\tilde{U}/\kappa$ by cooling from the initial cryogenic temperature
$T=0.1$ K. The blank regions correspond to the classical unstable
regime confirmed via the Routh-Hurwitz criterion. 
The black dash-dotted line indicates $\lambda(\omega_m)=0$.
The inset shows
the spectra of the input vacuum noise $S_{c}(\omega)$ (dashed), the
thermal noise $S_{Q}^{th}(\omega)$ (dotted), and sum of them (solid).
Other parameters are $(\kappa,g,\omega_{m},\gamma_{m})/2\pi$ = (1.5
}MHz, 0.35 Hz, 300 kHz, 0.1 Hz) and $\mathcal{P}=100$
mW.}
\end{figure}

\textit{Cryogenics} - We further examine the cooling of the membrane
when it is initially in a cryogenic environment with $T=0.1$ K, as
shown in Fig. \ref{fig:cryogenics}. Note that the phonon occupation
of the membrane can be of $10^{4}$ quanta even at a cryogenic temperature
of tens of millikelvin. In contrast to the case of room-temperature
environment, here the effect of the back-action force noise $S_{c}(\omega)$
can be stronger than that induced by the mechanical thermal noise
$S_{Q}^{th}(\omega)$. As a result, the overall noise spectrum is
frequency dependent, and the back-action force noise becomes the leading
limitation for effective cooling of the membrane, see the inset of
the Fig. \ref{fig:cryogenics}. Again, we find that in the bistable
regime $T_{{\rm eff}}$ gradually approaches the local minimum as
$s_{1}$ or $s_{2}$ gets close to the turning points, but the minimal
$T_{{\rm eff}}$ in the bistable region can be either at the upper
branch or lower branch. This can be understood again in terms of the optomechanical damping rate $(\gamma_{{\rm eff}}(\ensuremath{\omega_m})-\gamma_m)$ near the boundary with $\lambda(\omega_m)=0$ [as indicated by the dashed-dotted line], but one should note that the effective temperature becomes strongly dependent on the spectra of the
input vacuum noise in a cryogenic
environment, but not simply relevant to a large $\gamma_{{\rm eff}}(\ensuremath{\omega})$ as in the case of the room temperature. Moreover, we find that, in the monostable
regime, the minimum of the effective temperature is given by $T_{{\rm eff}}^{(\text{min})}\approx20.7$
$\mu$K with $U/\kappa=1.78\times 10^{-12}$ and $\Delta/\kappa=0.5$ (i.e. $\tilde{U}/\kappa=0.2$),
which corresponds to the quanta of phonon being less than one, and
is about two orders of magnitude less than the case of $U=0$. Despite
the parameter condition with respect to $T_{{\rm eff}}^{(\text{min})}$
is located at the monostable region (corresponding to the case of
only branch), one can reach ground state cooling under the unresolved sideband regime with $\kappa=2\pi\times1.5$ MHz, $\mathcal{P}=100$ mW, and $(\omega_{m}, \gamma_{m},g)/\kappa= (0.2,6.67\times10^{-8},2.33\times10^{-7})$.

\section{Further discussion and Conclusion}
In summary, we have studied mechanical cooling at the bistable regime of a dissipative optomechanical cavity with a large linewidth, which is modulated by the mechanical displacement. 
For appropriate laser driving frequencies, there exists a wide driving-power range on the order of 100 mW for observing optical bistability with $\kappa/2\pi=1.5$ MHz. The mechanical membrane can be optimally cooled down to 5 mK from the room temperature in the bistable region at the unresolved sideband regime, and can be cooled down close to the ground state from the cryogenic temperature $T=0.1$ K in the monostable region. For further studies, when the cavity linewidth is comparable to the mechanical frequency, one again finds the optimal $T_{\text{eff}}$ around $s_{1(2)}\sim0$ at the bistable regime with $T=293$ K. Moreover, increasing the laser driving power enables cooling of the membrane to a lower temperature for a weaker Kerr nonlinearity,
however, the laser detuning should be enlarged correspondingly in order to stabilize the system. Since the mechanical cooling strongly depends on the laser detuning and the Kerr nonlineartiy modified effective detuning, therefore, the lowest mechanical temperature that can be achieved will be limited by the free spectral range of
the cavity. Our result marks a crucial step for understanding the bistable behavior of the dissipative optomechanical systems, and has potential applications in preparation of nonclassical mechanical states and quantum information processing. 
\begin{acknowledgments}
H.W. acknowledges support from the National Natural Science Foundation
of China under Grants No. 11774058 and No. 12174058. Y.L. was supported
by the Natural Science Foundation of China (Grants No. 12074030 and
No. 12274107) and the Research Funds of Hainan University {[}Grant
No. KYQD(ZR)23010{]}.
\end{acknowledgments}

\bibliographystyle{apsrev4-1}
\bibliography{reference}

\end{document}